\documentclass[10pt,letterpaper]{article}
\usepackage[top=0.85in,left=2.75in,footskip=0.75in]{geometry}

\usepackage{latexsym}
\usepackage{graphicx}
\usepackage{amsthm}
\usepackage{makeidx}
\usepackage{authblk}
\usepackage{courier}
\usepackage{titletoc}

\newtheorem{property}{Property}
\newtheorem{definition}{Definition}

\usepackage{amsmath,amssymb}

\usepackage{changepage}

\usepackage[utf8x]{inputenc}

\usepackage{textcomp,marvosym}

\usepackage{cite}

\usepackage{nameref,hyperref}

\usepackage[right]{lineno}

\usepackage{microtype}
\DisableLigatures[f]{encoding = *, family = * }

\usepackage[table]{xcolor}

\usepackage{array}

\newcolumntype{+}{!{\vrule width 2pt}}

\newlength\savedwidth

\raggedright
\usepackage[aboveskip=1pt,labelfont=bf,labelsep=period,justification=raggedright,singlelinecheck=off]{caption}

\makeatletter
\renewcommand{\@biblabel}[1]{\quad#1.}
\makeatother

\usepackage{lastpage,fancyhdr,graphicx}
\usepackage{epstopdf}
\fancyheadoffset[L]{2.25in}
\fancyfootoffset[L]{2.25in}
\begin{document}
\vspace*{0.2in}

\begin{flushleft}
{\Large
\textbf\newline{Preemptive periodic epidemic control reduces life and healthcare system costs without aggravation of social and economic losses} 
}
\newline
\\
Giuseppe De Nicolao\textsuperscript{1*},
Marta Colaneri\textsuperscript{2\Yinyang},
Alessandro Di Filippo\textsuperscript{2\Yinyang},
Franco Blanchini\textsuperscript{3\ddag},
Paolo Bolzern\textsuperscript{4\ddag},
Patrizio Colaneri\textsuperscript{4\ddag},
Giulia Giordano\textsuperscript{5\ddag},
Raffaele Bruno\textsuperscript{2,6\ddag}
\\
\bigskip
\textbf{1} Department of Electrical, Computer and Biomedical Engineering, University of Pavia, Pavia, Italy
\\
\textbf{2} Division of Infectious Diseases I, Fondazione IRCCS Policlinico San Matteo, Pavia
\\
\textbf{3} Department of Mathematics, Computer Science and Physics, University of Udine, Udine, Italy
\\
\textbf{4} Department of Electronics, Information and Bioengineering, Politecnico di Milano, Milan, Italy
\\
\textbf{5} Department of Industrial Engineering, University of Trento, Trento, Italy
\\
\textbf{6} Department of Clinical, Surgical, Diagnostic, and Pediatric Sciences, University of Pavia, Pavia, Italy
\bigskip

%
%
\Yinyang These authors contributed equally to this work.

\ddag These authors also contributed equally to this work.




* giuseppe.denicolao@unipv.it

\end{flushleft}
\section*{Abstract}
Many countries are managing COVID-19 epidemic by switching between lighter and heavier restrictions. While an open-close and a close-open cycle have comparable socio-economic costs, the former leads to a much heavier burden in terms of deaths and pressure on the healthcare system. An empirical demonstration of the toll ensuing from procrastination was recently observed in Israel, where both cycles were enforced from late August to mid-December 2020, yielding some 1,600 deaths with open-close compared to 440 with close-open.

\section*{Author summary}
Two periodic epidemic control strategies are compared: (i) keep open and close only when the situation becomes too serious, (ii) close as soon as possible, and reopen only when the number of cases is low enough.  We demonstrate  that an anticipated action drastically reduces deaths and healthcare system costs without any significant increase in the other socio-economic costs.


\section*{Introduction}
%

Non Pharmacological Interventions (NPI) implemented  to contain the spread of SARS-CoV-2 are mainly concerned with the closure of production, commercial, educational and recreational activities, together with travel bans or restrictions.  These strict regulations have been effective in slowing down the rate of contagion and possibly anticipating the peak of infection cases \cite{Giordano_2020}, \cite{Prem_2020}. It is widely agreed that early timing in adopting restrictions is crucial \cite{Gerli_Minerva_2020}. However, due to the far-reaching social and economic consequences, the governments worldwide are facing an increasing pressure to ease restrictions.

In our analysis, we distinguish two broad categories of costs: health and human life costs of the epidemic on the one hand and socio-economic costs of NPIs on the other hand. The former ones include the impact of Covid-19 on deaths and the health status of the population and the healthcare system financial costs incurred to cure infected people. Examples of socio-economic cost of NPIs are the financial losses due to the restrictions imposed on industrial, commercial, and recreational activities, as well as the social costs due to the reduced effectiveness of distance education and the general deterioration of the quality of life.

Different policy responses are being taken by European governments to curb the spread of the virus \cite{FERRARESI2020109628}. Although with varying intensity, a common strategy in many countries has been to enforce harsher restrictions once the number of infected individuals rises to an alarming level, followed by a release when the number of infected has decreased and become manageable with the available public health resources \cite{normile2020suppress}. The closure trigger has been generally identified with the strain on healthcare systems, particularly intensive care units (ICUs). However, this widely-accepted strategy of closing only when approaching saturation of the ICU capacity may not be the most appropriate.  

Previous observational and modeling studies have suggested intermittent distancing measures to avoid critical care capacity overload in the absence of effective therapies and sustainable preventive measures for SARS-CoV-2 infection \cite{chowdhury2020dynamic}, \cite{ferguson2020impact},\cite{kissler2020projecting},\cite{bin2021post}.  Here, we consider a lock/release cycle driving the system to the same initial infection condition, possibly periodically repeated. Not only we show that the first-close-then-open approach has the potential to dramaticaly reduce fatality and ICU stress with respect to the miopic first-open-then-close strategy, but we also show that this comes at no additional socio-economic cost.

%
%
\section*{Results}
\subsection*{The average principle and its consequences on socio-economic costs}
An ICU-capacity approach can be modelled as an \emph{open-close} (OC) scheme: the reproduction number $R_t$ remains equal to its maximal value $r_o>1$ during a first \emph{open phase} of length $T_o$, followed by a \emph{close phase} of length $T_c$, characterized by $R_t$ constant and equal to its minimal value $r_c<1$. For the control strategy to be periodic, at the end of the whole cycle, lasting $T=T_o+T_c$ days, the number of active cases returns equal to the number $I_0$, observed at day zero.

By the Average principle (see Methods)  periodicity is achieved only if the reproduction number averages to one during the cycle, i.e. if 
$$\frac{r_oT_o+r_cT_c}{T}=1.$$
When the average principle is satisfied, the periodic application of the OC strategy yields a periodic profile of the active cases that returns to $I_0$ every $T$ days.

By just exchanging the open and close phases, another admissible periodic control strategy is obtained, named \emph{close-open} (CO). In particular, the reproduction number $R_t$ remains equal to $r_c<1$ during a first \emph{close phase} of length $T_c$, followed by an \emph{open phase} of length $T_o$, characterized by $R_t$ constant and equal to $r_o>1$.

According to the basic SIR model \cite{andersonreproduction},
$$
R_t=\frac{\beta(t)}{\gamma} \frac{S(t)}{N}
$$
where $1/\gamma$ is the average duration of infectiousness, $\beta(t)$ is the transmission parameter at time $t$, and $S(t)/N$ is the fraction of susceptible subjects $S(t)$ within the overall population of size $N$, consisting of susceptible, infected, and removed subjects. The transmission parameter $\beta(t)$ depends on biological, epidemiological, environmental and social factors and can be decreased by NPIs that reduce the rate at which individuals have close contact with other individuals. Therefore, a lower $\beta$ corresponds to larger socio-economic costs.
We assume that during a single period of the periodic control scheme, $S(t)/N$ does not undergo significant changes. This assumption is particularly plausible when letting the infection spread on a large scale would collapse the healthcare system. Then, $\beta(t)$ is approximately proportional to $R_t$, whose value essentially depends on the intensity of the NPIs. In particular, more restrictive measures lead to a decrease in $R_t$.

The proportionality between $R_t$ and $\beta(t)$ implies that the socio-economic costs of the OC and CO strategies are equivalent. In fact, in both cycles the time lengths of the open and close phases are identical, the only difference being the order. Possible differences may be due to seasonal economic activities, e.g. related to tourism, that are sensitive to translations of the bans. Nevertheless, for most activities and services, economic and social losses depend mainly on the duration and stringency of the restrictions enforced to keep $R_t$ within the desired limits. Having established the socio-economic equivalence of the two strategies, now it remains to compare their costs in health and human lives.

\subsection*{Periodic control strategies compared: open-close vs close-open}
A measure of the life and healthcare system costs of a periodic control strategy is given by deaths, hospital and ICU occupancies. In turn, these are proportional to the total number of new cases during one cycle, which equals the area under the curve (AUC) of the daily new cases during one cycle, see Methods. Under periodic control, this AUC is proportional to the AUC of the active cases that can therefore be used as a measure of the healthcare system costs (see Life and healthcare system costs in Methods). In particular, it offers a visually intuitive comparison of the costs of alternative strategies, see Fig. 1, where the AUCs are displayed of specular OC and CO cycles having closure and opening sub-intervals of the same length, but with the order exchanged.


\begin{figure}[h] 
\begin{center}
\includegraphics[scale=0.5] {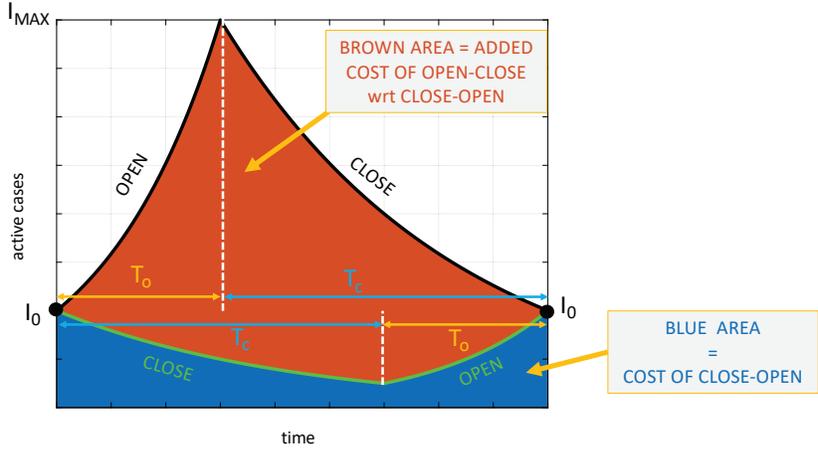}
\caption{Open-close (OC) vs close-open (CO). The AUCs are proportional to the life and healthcare system costs. The ratio of OC's cost to CO's one is equal to $I_{MAX}/I_0$.}
\label{fig_costs}
\end{center}
\end{figure}

As seen in Fig. 1, the life and healthcare system costs of the two strategies are dramatically different because returning to the initial number of active cases through the upper route, i.e.  with an exponential growth followed by an exponential decrease, yields a much greater AUC than following the lower route, consisting of an exponential decrease followed by an exponential growth. This kind of plot offers also a visually intuitive guideline to the decision maker dealing with periodic epidemic control. Indeed, it is of immediate understanding that to minimize its AUC you must keep the curve of active cases as low as you can, consistent with the constraint of returning to the starting value 

As shown in the Methods (Cost of open-close vs close-open), the ratio between the two AUCs obeys the formula
$$
\frac{\mathrm{AUC}_{OC}}{\mathrm{AUC}_{CO}}=K=\frac{I_{max}}{I_0},
$$
where $I_{max}$ denotes the maximum number of active cases reached during the OC cycle. For instance, if the number of active cases quadruples at the top of the open phase of OC, the associated life and healthcare system cost will be four times larger than the cost of the preemptive CO scheme.

The comparison of open-close and close-open control strategies is further illustrated in Panel A of Figure 2, displaying the simulated profile of active cases under an open-close cycle followed by a close-open one, each of 54 days. Both cycles consist of an open phase of duration $T_o= 31$ days and a close phase of duration $T_c=23$ days, but in reverse order. 

The profile of the active cases follows Equation (\ref{eq_OCprofile}) during the OC cycle and Equation (\ref{eq_COprofile})during the CO one, with $\alpha=0.0410$ day$^{-1}$ and $\beta=0.0553$ day$^{-1}$. Under these assumptions, at the end of each cycle, the active cases return to the initial value $I_0=21,000$, see points P, Q, and R. By Properties \ref{costOC} and \ref{costCO}, the predicted life and healthcare system costs of the two periodic strategies are proportional to the red and blue areas. The maximum number of active cases reached during the OC cycle is $I_{\mathrm{max}}=75,000$, so that  $I_{\mathrm{MAX}}/I_0=3.6$. Therefore, in this simulation the cost of the OC strategy is $3.6$ times larger than that of the CO strategy.

\begin{figure}[h] 
\begin{center}
\includegraphics[scale=0.37]{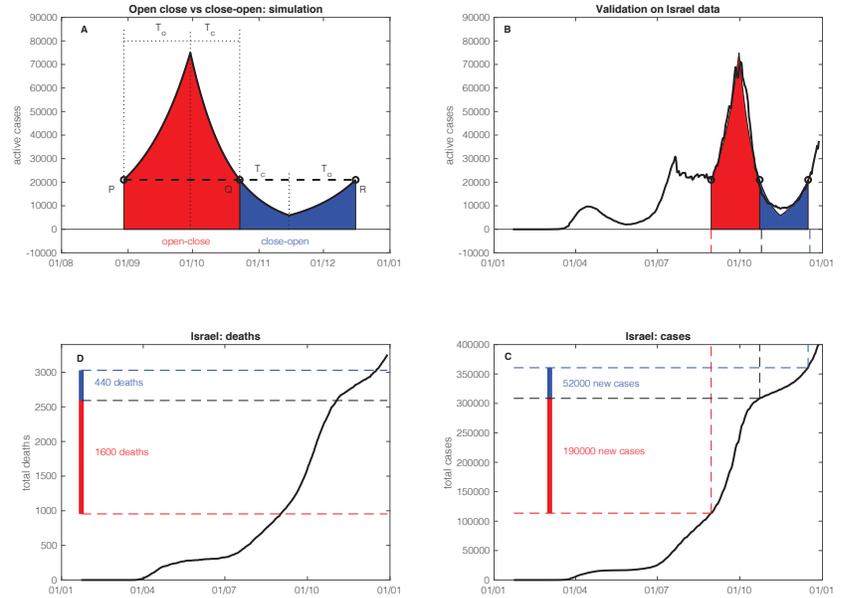}
\caption{Validation on Israel data. Periodic control strategies compared: in accordance with model predictions, a more than 3-fold death toll was observed in Israel after an open-close cycle followed by a close-open one.}
\label{fig_Israel_validation}
\end{center}
\end{figure}

\subsection*{Validation on Israel data}
The cost comparison presented in the previous section was empirically validated on Israel data retrieved from the COVID-19 Data Repository by the Center for Systems Science and Engineering at Johns Hopkins University \cite{dong2020interactive}.



During the 108-day period from August 30 to December 16, Israel's active cases (see panel B in Fig. 2) evolved very much like the simulation in panel A. 
In view of the similarity of Israel's active cases profile with a sequence of an ideal OC cycle followed by a CO one, the actual deaths can be used to validate our prediction (\ref{eq_costratio}) of the ratio between the OC and CO costs.

The first step is obtaining the total number of new cases occurring within the two cycles. For this purpose, it suffices to intersect the starting and ending dates of the cycles with the cumulative total cases, see panel C in Fig.2. It is found that there were some 190,000 cases during the OC cycle and 52,000 during the CO one

The final step is to assess the life costs by multiplying the cases by the estimated $\hat CFR =0.0085$, see Estimation of Israel's CFR in Methods. By properly scaling the axis of total deaths, this last conversion is visualized in Panel D of Fig. 2. It turns out that the OC and CO cycles caused some 1,600 and 440 deaths, respectively. The ratio is 3.7,  thus validating the formula (\ref{eq_costratio}) that predicted a 3.6-fold cost of the open-close strategy with respect to the close-open one.

The costs of hospital and ICU occupancy associated with the two strategies could be assessed in a similar way. Indeed, for a given ratio of infected subjects requiring hospitalization and ICU admission, these costs are again proportional to the new cases and hence to the AUC of the active cases.

It should be noted that even when the use of Equation (\ref{eq_costratio}) is precluded because changes in nonpharmaceutical interventions are too gradual to be approximated via ideal OC and CO cycles, the cost assessment based on the AUCs remains valid and is itself sufficient to show the advantage of preemptive strategies over procrastinating ones.

\section*{Discussion and Conclusion}

We compared two periodic epidemic control strategies: an open-close strategy (keep open and only close when the situation becomes too serious) and a close-open one (close as soon as possible, and reopen only when the number of cases is low enough).  The duration of the closure and opening periods being equal in both cases, the former strategy does not bring any socio-economic benefit, while the latter saves lives and reduces healthcare system costs.

An immediate and simple quantitative assessment of the cost of procrastinating intervention is given by the ratio $K$ between the maximum number of active cases during the open-close cycle and the initial number of active cases: the life and healthcare system costs of the procrastinating strategy are $K$ times larger than those of the preemptive one. In view of our analysis, it appears that an anticipated action is probably the most effective tool available to decision makers to drastically cut the life and healthcare system costs of the epidemic without a significant increase in the other socio-economic costs.

It is also interesting to note that a constant control strategy enforcing $R_t=1$, though preferable to the OC one, is outperformed by CO periodic control. In fact, it is reasonable to think that the overall socio-economic costs would depend on the average $R_t$, being therefore equal to the ones of the OC and CO strategies. On the other hand the health and human life costs of the constant strategy, measured by the AUC of the active cases, are clearly heavier than those of the CO strategy. Therefore, the epidemic is better controlled by a lock/release cycle rather than by a steady containment.

Our analysis assumes that the ratio $S(t)/N$ does not change significantly during a control period. If we consider its change, $R_t$, besides depending through $\beta$ on the stringency of the restrictions, is also proportional to the declining ratio of susceptible individuals. This means that, during the close phase of the OC strategy, $R_t=r_c$ can be enforced with NPIs that are less restrictive than those required to ensure $R_t=r_c$ during the close phase of the CO strategy, meaning that a periodic $R_t$ is no more associated with periodic NPIs. This observation, however, does not affect the validity of our conclusions for two reasons. Before the start of vaccination, the susceptible ratio cannot be allowed to grow quickly unless health structures are overwhelmed. Second, the gap between the health and human life costs of OC  and CO is so large to be robust in the face of this approximation.

Perhaps surprisingly, the socio-economic costs of the periodic strategies depend on the intensity and duration of NPIs, but are not affected by the initial number of infected individuals. This means that, once it has been decided to stabilize the epidemic by applying a periodic control, the socio-economic costs are the same whether we start with 100 or 100,000 infected people. In the latter case, however, the health and human life costs are 1,000 times larger: this constitutes a formidable argument in favor of a drastic abatement of the prevalence followed by a periodic CO strategy until  mass vaccination finally makes it possible to decrease $R_t$ without having to prolong stringent NPIs.

Our analysis does not accounts the effects of vaccination. Future work should address the quantitative assessment of the potential advantages of preemption during vaccination rollout. Although the analysis could somehow follow the tracks of the present paper, it is expected to be more complex, because the faster decline of the rate susceptible individuals prevents a simple use of a periodic paradigm.

\section*{Methods}
\subsection*{Notation}
\begin{description}

\item
$t \in [0, T]$: time (days)

\item
$I(t)$: stock of infected subjects at time $t$ also called \emph{active cases}.

\item
$n(t)$: daily rate of new infected cases at time $t$.

\item
$\gamma$: removal rate of infected subjects.

\item
$R_t$: reproduction number at time $t$.

\item
$S_R(t) = \int_0^{t}R_{\tau} d \tau$: Area Under the Curve (AUC) of the reproduction number.

\item
$r_c <1$ minimal $R_t$ achieved with stronger nonpharmaceutical interventions.

\item
$r_o>1$: maximal value of $R_t$, achieved with milder nonpharmaceutical interventions.

\item \emph{Open-close (OC) cycle}: the adoption of nonpharmaceutical interventions such that 
$$
R_t= r_o,\quad  0 \le t < T_o, \quad R_t= r_c,\quad T_o \le t < T_o+T_c
$$

\item \emph{Close-open (CO) cycle}: the adoption of nonpharmaceutical interventions such that 
$$
R_t= r_c,\quad  0 \le t < T_c, \quad R_t= r_o,\quad T_c \le t < T_c+T_o
$$

\end{description}


\subsection*{Balance equation} \label{S2 Balance equation}
The stock of infected is described as a single compartment obeying the balance equation
\begin{equation} \label{balance}
\dot I(t) = - \gamma I(t) + n(t), \qquad I(0)=I_0
\end{equation}
Assuming that $n(t)=R_t \gamma I(t)$, it follows that
\begin{equation} \label{eq_infect}
\dot I(t) = (R_t -1) \gamma I(t), \qquad I(0)=I_0
\end{equation}
whose solution is
\begin{equation} \label{eq_sol_infect}
I(t) = I_0 e^{\gamma \int_0^{t}(R_{\tau}-1) d \tau} = I_0 e^{\gamma(S_R(t) - t)}
\end{equation}


\subsection*{Periodic epidemic control: average principle}
\label{S3 Periodic epidemic control: average principle}
\begin{definition}
$R_t$ is said to achieve periodic epidemic control of period $T$ if $I(T)=I_0$. 
\end{definition}

In view of (\ref{eq_sol_infect}), $I(T)=I_0$ if and only if $S_R(T)=T$, so that the following property holds.

\begin{property} 
If $R_t$ achieves periodic control of period $T$, then it satisfies the \emph{average principle}
$$
\frac{S_R(T)}{T} = 1
$$
i.e. the average reproduction number $R_t$ in the interval $[0, T]$ is equal to one.
\end{property}

Letting $T=T_o+T_c$, for both the OC and the CO cycle it holds that
\begin{equation} \label{eq_avg}
S_R(T) = r_o T_o + r_c T_c
\end{equation}
Therefore, the average principle implies that a cycle (either OC or CO) achieves periodic control of period $T=T_o+T_c$ if
$$
\frac{r_o T_o + r_c T_c}{T}=1
$$
This means that, under periodic control with a given period $T$, the lengths  $T_o$ and $T_c$ of the open and close phases depend only on $r_o$ and $r_c$, regardless of the order in which they are applied. In particular,
$$
T_o=\frac{1-r_c}{r_o-r_c} T, \quad T_c=\frac{r_o-1}{r_o-r_c} T
$$

\subsection*{Life and healthcare system costs}
\label{sect_AUCcosts}

The life and healthcare system costs are obviously proportional to the total number of cases, i.e. to the AUC of $n(t)$ over the considered time interval $T$. In this section, it is shown that under $T$-periodic control they are also proportional to the AUC of the active cases $I(t)$ over the period $T$. This property will be instrumental in obtaining an immediate comparison of OC vs CO in terms of the maximum number of active cases reached during the OC cycle.

Let $a(t)$ denote the solution of (\ref{balance}) with $I_0=0$ and define
$$
\mathrm{AUC}_f(t_1,t_2)= \int_{t_1}^{t_2}f(t) dt, \quad \mathrm{AUC}_f= \int_0^{\infty}f(t) dt
$$

\begin{property}
$\mathrm{AUC}_n = \gamma \mathrm{AUC}_a$.
\end{property}

\begin{proof}
In view of (\ref{balance}),
$$
a(t) =\int_0^t e^{-\gamma(t-\tau)} n(\tau) d\tau
$$
Then,
\begin{eqnarray*}
\mathrm{AUC}_a &=& \int_0^{\infty} \int_0^t e^{-\gamma(t-\tau)} n(\tau) d\tau dt = \int_0^{\infty} n(\tau) \int_{\tau}^{\infty} e^{-\gamma(t-\tau)}  dt \, d\tau \\
&=& \int_0^{\infty} n(\tau) \int_0^{\infty} e^{-\gamma s}  ds \, d\tau = \mathrm{AUC}_n/\gamma
\end{eqnarray*}
\end{proof}

\begin{property}
$\mathrm{AUC}_n(0,T) = \gamma \mathrm{AUC}_a(0,T) + a(T) $.
\end{property}

\begin{proof}
Define
$$
\tilde n(t)= \left\{\begin{array}{cc}n(t), & 0\le T \\0 & t>T\end{array}\right.
$$
and let $\tilde a(t), t \ge 0$ be the solution of (\ref{balance}) with $I_0=0$ and $n(t)=\tilde n(t)$.
Then, 
$$
\mathrm{AUC}_n(0,T) = \mathrm{AUC}_{\tilde n} = \gamma \mathrm{AUC}_{\tilde a} = \gamma \mathrm{AUC}_a(0,T) + \gamma \mathrm{AUC}_{\tilde a}(T,\infty)
$$
Observing that $\tilde a(t)=a(T) e^{-\gamma t}, t > T$,
$$
\mathrm{AUC}_{\tilde a}(T,\infty) = a(T) \int_T^{\infty}  e^{-\gamma t} dt= \frac{a(T)}{ \gamma}
$$
and the thesis follows.
\end{proof}

\begin{property}
For a given $I_0$, let $n(t)$ be such that $I(T)=I_0$. Then,
$$
\mathrm{AUC}_n(0,T) = \gamma \mathrm{AUC}_I(0,T)
$$
\end{property}

\begin{proof}
Since
$
I(t) = e^{-\gamma t} I_0 + a(t)
$,
the periodicity constraint $I(T)=I_0$ implies that
$$
a(T) = (1 - e^{-\gamma T}) I_0
$$
Then,
\begin{eqnarray*}
\mathrm{AUC}_I(0,T)&=&I_0 \int_0^T e^{-\gamma t} dt + \mathrm{AUC}_a(0,T) \\
&=&\frac{I_0}{\gamma} (1 - e^{-\gamma T}) + \mathrm{AUC}_a(0,T)\\
&=& \frac{a(T)}{\gamma} + \mathrm{AUC}_a(0,T) \\
&=& \frac{\mathrm{AUC}_n(0,T)}{\gamma}
\end{eqnarray*}
thus proving the thesis.
\end{proof}

Since the life and healthcare system costs of different periodic control strategies are proportional to the AUCs of the active cases, $C=\mathrm{AUC}_I(0,T)$ can be used as a measure of such costs.


\subsection*{Cost of open-close vs close-open}
\label{S5 Cost of open-close vs close-open}

Let $\alpha=\gamma(r_o-1)$ and $\beta=\gamma(1-r_c)$. Moreover, we denote by $I_{\mathrm{MAX}}$ the maximum number of active cases reached at the end of the open phase of the OC cycle:
\begin{equation} \label{eq_Imax}
I_{\mathrm{MAX}} = I_0 e^{\alpha T_o} 
\end{equation}

\begin{property} \label{costOC}
The cost $C_{oc}$ of the periodic OC strategy is
\begin{equation} \label{eq_cost_oc}
C_{oc}= \left( \frac{1}{\beta}+\frac{1}{\alpha} \right) \left(e^{\alpha T_o}- 1 \right)I_0 = \left( \frac{1}{\beta}+\frac{1}{\alpha} \right) \left( I_{\mathrm{MAX}} - I_0 \right) 
\end{equation}
where
\begin{equation} \label{eq_To}
T_o = \frac {\beta T}{\alpha + \beta}
\end{equation}
\end{property}

\begin{proof}
From (\ref{eq_avg}) it follows that
$$
\alpha T_o - \beta T_c = 0
$$
which in turn implies
$$
T_o = \frac {\beta T}{\alpha + \beta}
$$
From (\ref{eq_infect}) it follows that
\begin{equation} \label{eq_OCprofile}
I(t) = \left\{\begin{array}{cc} I_0 e^{\alpha t} ,& 0\le t \le T_o \\I_0 e^{\alpha T_o}e^{-\beta (t-T_o)}, & T_o \le t \le T \end{array}\right.
\end{equation}
Therefore, observing that $\exp(\alpha T_o) \exp(- \beta (T-T_o)) = 1 $,
\begin{eqnarray*}
C_{oc} &=&\mathrm{AUC}_I(0,T)= \int_0^{T_o}I_0 e^{\alpha t}dt + \int_{T_o}^{T}I_0 e^{\alpha T_o}e^{ -\beta(t-T_o)} dt  \\
 &=& \frac{I_0}{\alpha} \left( e^{\alpha T_o}  - 1  \right) + \frac{I_0 e^{\alpha T_o} }{\beta} \left( 1 - e^{\beta(T- T_o)}   \right)   \\
 &=& I_0 \left( \frac{1}{\beta}+\frac{1}{\alpha} \right) \left( e^{\alpha T_o} - 1 \right) 
\end{eqnarray*}
and the thesis follows.
\end{proof}

\begin{property} \label{costCO}
The cost $C_{co}$ of the periodic CO strategy is
\begin{equation} \label{eq_cost_co}
C_{co}= \left( \frac{1}{\beta}+\frac{1}{\alpha} \right) \left( 1 -  e^{- \alpha T_o} \right)I_0 =e^{- \alpha T_o} \left( \frac{1}{\beta}+\frac{1}{\alpha} \right) \left(  I_{\mathrm{MAX}} -I_0  \right) 
\end{equation}
\end{property}

\begin{proof}
From (\ref{eq_infect}) it follows that
\begin{equation}  \label{eq_COprofile}
I(t) = \left\{\begin{array}{cc} I_0 e^{-\beta t} ,& 0\le t \le T_c \\I_0 e^{-\beta T_c}e^{\alpha (t-T_c)}, & T_c \le t \le T \end{array}\right.
\end{equation} 
Therefore, observing that $\exp(-\beta T_c) \exp(\alpha (T-T_c)) = 1 $,
\begin{eqnarray*}
C_{co}&=&\mathrm{AUC}_I(0,T)= \int_0^{T_c}I_0 e^{-\beta t}dt + \int_{T_c}^{T}I_0 e^{-\beta T_c}e^{\alpha (t-T_c)} dt  \\
 &=& \frac{I_0}{\beta} \left( 1 - e^{-\beta T_c}  \right) + \frac{I_0 e^{-\beta T_c}}{\alpha} \left( e^{\alpha(T- T_c)}  - 1  \right)   \\
 &=& I_0 \left( \frac{1}{\beta}+\frac{1}{\alpha} \right) \left( 1 - e^{-\beta T_c} \right) \\
 &=& e^{- \alpha (T-T_c)} \left( \frac{1}{\beta}+\frac{1}{\alpha} \right) \left( e^{ \alpha (T-T_c)}  - 1 \right) I_0 \\
 &=& e^{- \alpha T_o} \left( \frac{1}{\beta}+\frac{1}{\alpha} \right) \left( e^{ \alpha T_o}I_0  - I_0 \right)  \\
\end{eqnarray*}
and the thesis follows.
\end{proof}

\begin{property}
The cost $C_{co}$ of the CO strategy is less then the cost $C_{\mathrm{const}}$ of a constant control strategy maintaining $I(t)=I_0, 0\le t \le T$:
$$
C_{co} < C_{\mathrm{const}} = I_0 T
$$
\end{property}

\begin{proof}
Let
$$
\rho=\left( \frac{1}{\beta}+\frac{1}{\alpha} \right)^{-1}
$$
From (\ref{eq_To}) it follows that $\alpha T_o = \rho T$.
Equation (\ref{eq_cost_co}) can be rewritten as
\begin{equation} 
C_{co}= \frac{I_0}{\rho} \left(1 - e^{-\alpha T_o} \right) = \frac{I_0}{\rho} \left(1 - e^{-\rho T} \right) 
\end{equation}
Then, letting $y=\rho T$, we have that
$$
\frac{C_{co}}{C_{\mathrm{const}} } = \frac{1-e^{-y}}{y}
$$
The right-hand side is always strictly less than one, $\forall y > 0$, thus proving the thesis.
\end{proof}

\begin{property} \label{costs}
The ratio between the costs of the periodic open-close and close-open strategies is equal to the maximum growth ratio of the active cases during the open-close cycle:
\begin{equation} \label{eq_costratio}
\frac{C_{oc}}{C_{co} } =\frac{ I_{\mathrm{MAX}} }{I_0}
\end{equation}
\end{property}

\begin{proof}
In view of (\ref{eq_cost_oc}) and (\ref{eq_cost_co}), the proof follows directly from (\ref{eq_Imax}).
\end{proof}



\subsection*{Israel data}
\label{S6 Israel data}
\begin{figure}[h] 
\begin{center}
\includegraphics[scale=0.65]{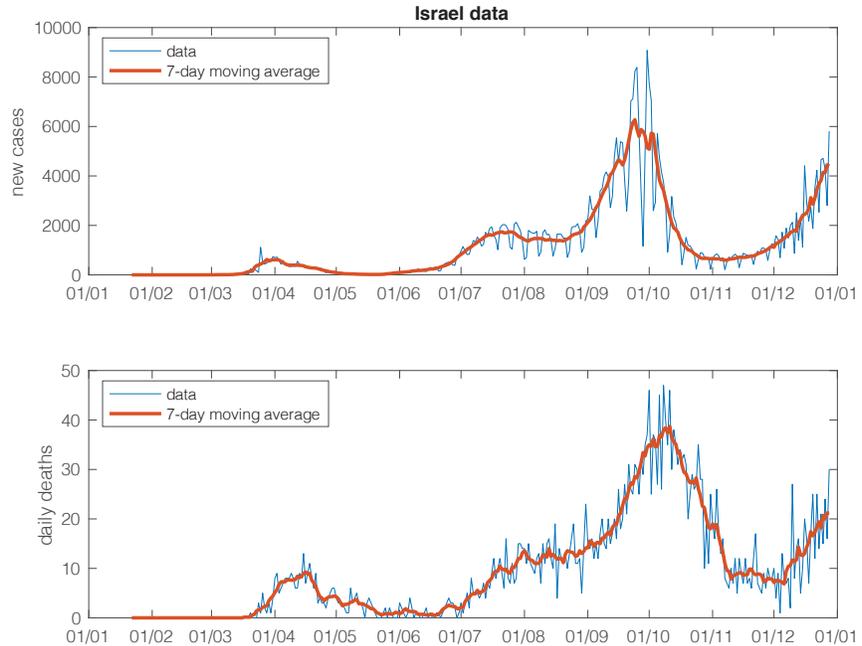}
\caption{Israel data: time series of new cases (top) and daily deaths (bottom).}
\label{fig_Israel_new_deaths}
\end{center}
\end{figure}

The data were downloaded from the COVID-19 Data Repository by the Center for Systems Science and Engineering at Johns Hopkins University \cite{dong2020interactive}. The downloaded time series included: deaths, confirmed cases, and recovered ones. From these cumulative data, daily deaths and new cases, displayed in Fig. \ref{fig_Israel_new_deaths}, were obtained by differencing consecutive data. Active cases were computed by subtracting deaths and recovered cases from the confirmed ones.

\begin{figure}[h]
\begin{center}
\includegraphics[scale=0.5]{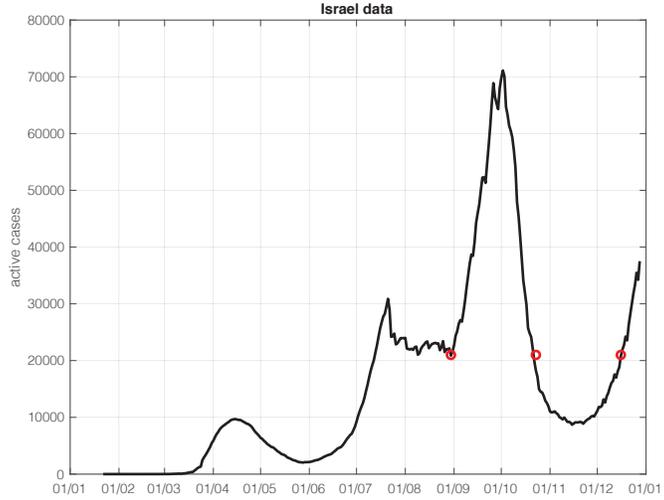}
\caption{Israel data: active cases. Circles, spaced 54 days apart on the time axis starting from August 30, are placed in correspondence of 21,000 cases.}
\label{fig_Israel_active}
\end{center}
\end{figure}

By inspecting the active cases, see Fig. \ref{fig_Israel_active}, it is seen that on August 30 they were equal to 20,876. Starting from that date, they had an upswing, reaching 71,114 on October 3, when a downswing started, leading to a nadir of 8,697 cases on November 16. On December 16, the active cases had risen to 20,791, close to those of August 30. In other words, after an upper arc and a lower one, each during about 54 days, the active cases returned to their initial value.

Taking into account a physiological delay between interventions and published data, the time course of the active cases can be associated to the nonpharmaceutical interventions. Schools reopened on September 1, following the Education Ministry's Safe Learning Plan that provided for smaller class sizes in the younger grades and at least two days in school from grades 5 through 12. Due to the growth of the cases, the Government resolved to impose a total national lockdown effective from Friday September 25 for the duration of 14 days, reserving the option to extend it as it actually happened. In the first phase of exit, which came into effect on 18 October, beaches, nature reserves and parks in Israel were reopened, as were commercial activities without public reception and take-away services. Starting from November 1, the lockdown was gradually eased. For instance, pupils of grades 1-4 were readmitted to school and on November 8 shops reopened with 4-customers only limit. Relaxation of restrictions in most of the country was followed by a growth of the cases that has provoked a new national lockdown to begin on 27 December 2020.

\subsection*{Estimation of Israel's CFR}
\label{S7 Estimation of Israel's CFR}
In order to assess the life cost of new cases, an estimate of the CFR (Case Fatality Rate) is needed. This parameter is affected by the testing protocol and healthcare system reaction. Therefore, it was estimated using only data posterior to June 1, 2020, so as to discard first-wave data.

It is assumed that the daily deaths $d(t)$ depend on the previous daily new cases $n(t)$ according to the equation
\begin{equation} \label{eq_death_convol}
d(t) = \sum_{i=0}^{\infty} w(i) n(t-i)
\end{equation}
where the weight $w(i)$ denotes the fraction of infected subjects at time $t-i$ that die at time $t$. The Case Fatality Rate is then given by
$$
\mathrm{CFR} = \sum_{i=0}^{\infty} w(i)
$$
In order to estimate the weights, a simple exponential model with delay was assumed:
$$
w(i)=\left\{\begin{array}{cc}0, & i <k \\b a^{i-k}, & i \ge k \end{array}\right.
$$
where $k$ is the delay and $a$ and $b$ are unknown parameters.
\begin{figure}[h]
\begin{center}
\includegraphics[scale=0.4]{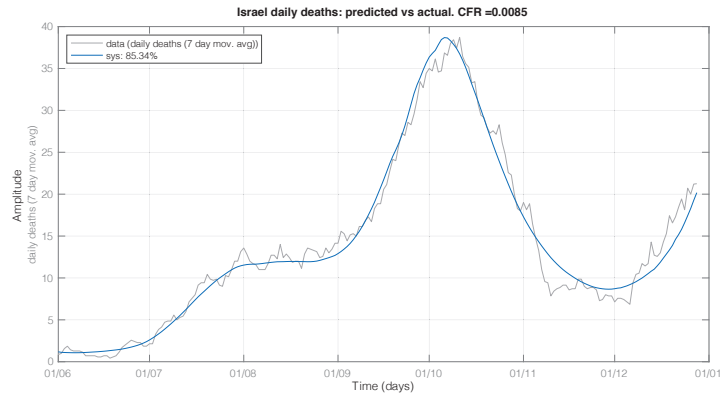}
\caption{Daily deaths predicted by weighted average of new cases, with delay equal to 3 and exponentially decaying weights.}
\label{fig_Israel_CFR}
\end{center}
\end{figure}
As seen in Fig. \ref{fig_Israel_new_deaths}, both the new cases and the daily deaths exhibit an apparent weekly seasonality. In the estimation procedure, the original series  were therefore replaced by their 7-day moving average (red lines in Fig. \ref{fig_Israel_new_deaths}).

For given values of $a,b,k$, Equation (\ref{eq_death_convol}) can be used to predict daily deaths. 

Israel data from June 1 to December 29 were used to estimate the parameters via least squares using the function \texttt{oe.m} of the MATLAB System Identification Toolbox, version R202b. The estimation of $a$ and $b$ was repeated for all delays $k$ ranging from $0$ to $15$ and the delay $k=3$ associated with the best sum of squares was eventually selected.
In spite of the simplicity of the exponential-with-delay model, the time series of deaths is explained  in a very satisfactory way, see Fig. \ref{fig_Israel_CFR}.

The estimated parameters and their percent coefficients of variation were:
$$
\left.\begin{array}{lll}\hat a = & 0.943, & CV(\%) = 0.34 \\ \hat b=  &0.000485, &CV(\%)= 5.17\end{array}\right.
$$
Then, recalling the formula for the sum of the harmonic series,
$$
\hat CFR= \sum_{i=0}^{\infty} \hat w(i) = \frac{\hat b}{1-\hat a} = 0.0085
$$
The estimated $\hat CFR$ is in good agreement with the value 0.008, currently (30.12.2020) reported in the COVID-19 Data Repository at Johns Hopkins University \cite{dong2020interactive}.


\section*{Acknowledgments}
This work has been partially supported by the Italian Ministry for Research in the framework of the 2017 Program for Research Projects of National Interest (PRIN), Grant no. 2017YKXYXJ.

\nolinenumbers

%
%
%

\bibliography{PLOS.bib}

%
%


\end{document}